\documentclass[preprint,aps,prd,nofootinbib,preprintnumbers,amsmath,amssymb]{revtex4}

\newcommand\fft[2]{\frac{#1}{#2}}
\newcommand\ft[2]{{\textstyle\frac{#1}{#2}}}
\newcommand\nn{{\nonumber}}

\begin{document}
\preprint{MCTP-10-14}

\title{Consistent massive truncations of IIB supergravity on Sasaki-Einstein
manifolds}

\author{James T.~Liu}
\email{jimliu@umich.edu}
\affiliation{Michigan Center for Theoretical Physics,
Randall Laboratory of Physics,
The University of Michigan,
Ann Arbor, MI 48109--1040, USA}

\author{Phillip Szepietowski}
\email{pszepiet@umich.edu}
\affiliation{Michigan Center for Theoretical Physics,
Randall Laboratory of Physics,
The University of Michigan,
Ann Arbor, MI 48109--1040, USA}

\author{Zhichen Zhao}
\email{zhichen@umich.edu}
\affiliation{Michigan Center for Theoretical Physics,
Randall Laboratory of Physics,
The University of Michigan,
Ann Arbor, MI 48109--1040, USA}

\begin{abstract}
Recent work on holographic superconductivity and gravitational duals of
systems with non-relativistic conformal symmetry have made use of consistent
truncations of $D=10$ and $D=11$ supergravity retaining some massive modes
in the Kaluza-Klein tower.  In this paper we focus on reductions of IIB
supergravity to five dimensions on a Sasaki-Einstein manifold, and extend
these previous truncations to encompass the entire bosonic sector of
gauged $D=5$, $\mathcal N=2$ supergravity coupled to massive multiplets up
to the second Kaluza-Klein level.  We conjecture that a necessary condition
for the consistency of massive truncations is to only retain the lowest modes
in the massive trajectories of the Kaluza-Klein mode decomposition of the
original fields.  This is an extension of the well-known result that
consistent truncations may be obtained by restricting to the singlet sector
of the internal symmetry group.
\end{abstract}

\maketitle

\section{Introduction}

Recent developments in AdS/CFT have expanded the scope of applications
from the realm of strongly coupled relativistic gauge theories to various
condensed matter systems whose dynamics are expected to be described by
a strongly coupled theory.  These include systems with behavior governed by
a quantum critical point \cite{Herzog:2007ij,Hartnoll:2007ih}, as well
as cold atoms and similar systems exhibiting non-relativistic conformal
symmetry \cite{Son:2008ye,Balasubramanian:2008dm}.  Much current attention
is also directed towards holographic descriptions of superfluids and
superconductors
\cite{Gubser:2008px,Hartnoll:2008vx,Herzog:2008he,Hartnoll:2008kx}.

The main feature used in the construction of a dual model of superconductivity
is the existence of a charged scalar field in the dual AdS background
\cite{Hartnoll:2008vx,Hartnoll:2008kx}.  Turning
on temperature and non-zero chemical potential corresponds to working with
a charged black hole in AdS.  Then, as the temperature is lowered, the charged
scalar develops an instability and condenses, so that the black hole develops
scalar hair%
\footnote{Recent models have generalized this construction to encompass
both p-wave \cite{Gubser:2008wv,Roberts:2008ns} and d-wave \cite{Chen:2010mk}
condensates.}.
This condensate breaks the U(1) symmetry, and is a sign of
superconductivity (in the case where the U(1) is ``weakly gauged'' on the
boundary).

The basic model dual to a 2+1 dimensional superconductor is simply that
of a charged scalar coupled to a Maxwell field and gravity, and may be
described by a Lagrangian of the form
\begin{equation}
\mathcal L_4=R+\fft6{L^2}-\fft14F_{\mu\nu}^2-|\partial_\mu\psi-iqA_\mu\psi|^2
-m^2|\psi|^2.
\end{equation}
The properties of the system may then be studied for various values of mass
$m$ and charge $q$.  While this is a perfectly acceptable framework, a
more complete understanding demands that this somewhat phenomenological
Lagrangian be embedded in a more complete theory such as string theory, or at
least its supergravity limit.  For AdS$_4$ duals of 2+1 dimensional
superconductors, this was examined at the linearized level in
\cite{Denef:2009tp}, and embedded into $D=11$ supergravity at the full
non-linear level in \cite{Gauntlett:2009zw,Gauntlett:2009dn,Gauntlett:2009bh}
for the case $m^2L^2=-2$ and $q=2$.  Similarly, a IIB supergravity model for
an AdS$_5$ dual to 3+1 dimensional superconductors was constructed in
\cite{Gubser:2009qm} with $m^2L^2=-3$ and $q=2$.

The AdS$_4$ model of \cite{Gauntlett:2009zw,Gauntlett:2009dn,Gauntlett:2009bh}
and the AdS$_5$ model of \cite{Gubser:2009qm} are based on Kaluza-Klein
truncations on squashed Sasaki-Einstein manifolds.  They both have the unusual
feature where the $q=2$ charged scalar arises from the massive level of the
Kaluza-Klein truncation.  This appears to go against the standard lore of
consistent truncations, where it was thought that truncations keeping only
a finite number of massive modes would necessarily be inconsistent.  A
heuristic argument is that states in the Kaluza-Klein tower carry charges
under the internal symmetry, and hence would couple at the non-linear level
to source higher and higher states, all the way up the Kaluza-Klein tower.
This hints that one way to obtain a consistent truncation is simply to truncate
to singlets of the internal symmetry group, and indeed such a construction
is consistent.  An example of this is a standard torus reduction, where only
zero modes on the torus are kept.  On the other hand, sphere reductions to
maximal gauged supergravities in $D=4$, 5 and 7 do not follow this rule, as
they are expected to be consistent, even though some of the lower-dimensional
fields (such as the non-abelian graviphotons) are charged under the
$R$-symmetry.  In fact, the issue of Kaluza-Klein consistency is not yet
fully resolved, and often must be treated on a case by base basis.  This
has led us to explore the squashed Sasaki-Einstein compactifications to see
if additional consistent massive truncations may be found.

In addition to embedding holographic models of superconductivity into
string theory, several groups have demonstrated the embedding of dual
non-relativistic CFT backgrounds into string theory
\cite{Herzog:2008wg,Maldacena:2008wh,Adams:2008wt}.  These geometries
where originally constructed from a toy model of a massive vector
field coupled to gravity with a negative cosmological constant
\cite{Son:2008ye,Balasubramanian:2008dm} of the form (given here for
a deformation of AdS$_5$):
\begin{equation}
\mathcal L_5=R+\fft{12}{L^2}-\fft14F_{\mu\nu}^2-\fft{m^2}2A_\mu^2,
\end{equation}
with mass related to the scaling exponent $z$ according to $m^2L^2=z(z+2)$.
The $z=2$ and $z=4$ models ($m^2L^2=8$ and $m^2L^2=24$, respectively) were
subsequently realized within IIB supergravity in terms of consistent
truncations retaining a massive vector (along with possibly other fields
as well) \cite{Herzog:2008wg,Maldacena:2008wh,Adams:2008wt}.  These results
have further opened up the possibility of obtaining large classes of
consistent truncations retaining massive modes of various spin.

\subsection{Consistent massive truncations of IIB supergravity}

For the most part, the massive consistent truncations used in the
study of AdS/condensed matter systems have not been supersymmetric%
\footnote{The massive truncation given in \cite{Gauntlett:2009zw}
is supersymmetric, although the connection to a holographic superconductor
was done through the non-supersymmetric skew-whiffed case.}.
Nevertheless this has motivated us to investigate the possibility of
obtaining new supersymmetric massive truncations of IIB supergravity.
In particular, we are mainly interested in reducing IIB supergravity on
a Sasaki-Einstein manifold to obtain gauged supergravity in $D=5$
coupled to possibly massive supermultiplets.

Following the construction of $D=11$ supergravity \cite{Cremmer:1978km}
and the realization that it admits an AdS$_4\times S^7$ vacuum solution
\cite{Freund:1980xh}, it was soon postulated that the Kaluza-Klein reduction
on the sphere would give rise to gauged $\mathcal N=8$ supergravity at
the ``massless'' Kaluza-Klein level \cite{Duff:1982gj,Duff:1982yw,Duff:1983gq}.
This notion was reinforced by a linearized Kaluza-Klein mode analysis
demonstrating that the full spectrum of Kaluza-Klein excitations falls into
supermultiplets of the $D=4$, $\mathcal N=8$ superalgebra $\mathrm{OSp}(4|8)$
\cite{Biran:1983iy,Sezgin:1983ik,Casher:1984ym}.  However, demonstrating
full consistency of the non-linear reduction to gauged $\mathcal N=8$
supergravity has remained elusive.  Nevertheless, all indications are that
the reduction is consistent \cite{de Wit:1986iy}, and this has in fact been
demonstrated for the related case of reducing to $D=7$ on $S^4$
\cite{Nastase:1999cb,Nastase:1999kf}.

The story is similar for the case of IIB supergravity reduced on $S^5$.
A linearized Kaluza-Klein mode analysis demonstrates that the spectrum of
Kaluza-Klein excitations falls into complete supermultiplets of the $D=5$,
$\mathcal N=8$ superalgebra $\mathrm{SU}(2,2|4)$, with the lowest one
corresponding to the ordinary $\mathcal N=8$ supergravity multiplet
\cite{Gunaydin:1984fk,Kim:1985ez}.  In this case, only partial results are
known about the full non-linear reduction to gauged supergravity, but there
is strong evidence for its consistency
\cite{Khavaev:1998fb,Cvetic:1999xp,Lu:1999bw,Cvetic:2000nc}.

More generally, it was conjectured in \cite{Pope:1985jg,Duff:1985jd} and
\cite{Gauntlett:2007ma}, that, for any supergravity reduction, it is always
possible to consistently truncate to the supermultiplet containing the
massless graviton.  This is a non-trivial statement, as the truncation must
satisfy rather restrictive consistency conditions related to the gauge
symmetries generated by the isometries of the internal manifold
\cite{Duff:1984hn,Hoxha:2000jf}.  This conjecture has recently been shown to
be true for Sasaki-Einstein reductions of IIB supergravity on $SE_5$
\cite{Buchel:2006gb} and $D=11$ supergravity on $SE_7$ \cite{Gauntlett:2007ma},
yielding minimal $D=5$, $\mathcal N=2$ and $D=4$, $\mathcal N=2$ gauged
supergravity, respectively (see also \cite{Gauntlett:2006ai,Gauntlett:2007sm}).

While states in the same supermultiplet do not necessarily have the same mass
in gauged supergravity, the minimal supergravity multiplets, which contain
the graviton, gravitino and a graviphoton, are in fact massless.  Thus one
may suspect that truncations to massless supermultiplets are necessarily
consistent.  However, it turns out that this is not the case.  This was
explicitly demonstrated in \cite{Hoxha:2000jf}, where, for example, it was
shown to be inconsistent to retain the $\mathrm{SU}(2)\times\mathrm{SU}(2)$
vector multiplets
that naturally arise in the compactification of IIB supergravity on $T^{1,1}$.

For many of the above reasons, it has often been a challenge to explore
consistent supersymmetric truncations, even at the massless Kaluza-Klein
level.  However, bosonic truncations retaining massive breathing and
squashing modes \cite{Bremer:1998zp}  have been known to be consistent for
some time.  In this case, consistency is guaranteed by retaining only
singlets under the internal symmetry group $\mathrm{SU}(4)\times\mathrm{U}(1)$
for the squashed $S^7$ or $\mathrm{SU}(3)\times\mathrm{U}(1)$ for the squashed
$S^5$.  The
supersymmetry of background solutions involving the breathing and squashing
modes was explored in \cite{Liu:2000gk}, where it was further conjectured
that a supersymmetric consistent truncation could be found that retains
the full breathing/squashing supermultiplet.

Although this massive consistent truncation conjecture was made for squashed
sphere compactifications, it naturally generalizes to compactification on
more general internal spaces, such as Sasaki-Einstein spaces.  For $D=11$
supergravity compactified on a squashed $S^7$, written as U(1) bundled
over $CP^3$, truncation of the $\mathcal N=8$ Kaluza-Klein spectrum to
SU(4) singlets under the decomposition $\mathrm{SO}(8)\supset\mathrm{SU}(4)
\times\mathrm{U}(1)$ yields the $\mathcal N=2$ supergravity multiplet%
\footnote{The $\mathrm{OSp}(4|2)$ super-representations $\mathcal D(E_0,s)_q$
and SO(2,3) representations $D(E_0,s)_q$ are labeled
by energy $E_0$, spin $s$ and U(1) charge $q$ under $\mathrm{OSp}(4|2)\supset
\mathrm{SO}(2,3)\times\mathrm{U}(1)\supset\mathrm{SO}(2)\times\mathrm{SO}(3)
\times\mathrm{U}(1)$.}
\begin{equation}
n=0:\qquad\mathcal D(2,1)_0=D(3,2)_0+D(\ft52,\ft32)_{-1}+D(\ft52,\ft32)_1
+D(2,1)_0,
\label{eq:n=0}
\end{equation}
at the massless ($n=0$) Kaluza-Klein level.  No SU(4) singlets survive
at the first ($n=1$) massive Kaluza-Klein level, and the breathing and
squashing modes finally make their appearance at the second ($n=2$)
Kaluza-Klein level in a massive vector multiplet \cite{Liu:2000gk}
\begin{eqnarray}
n=2:\qquad\mathcal D(4,0)_0&=&D(5,1)_0+D(\ft92,\ft12)_{-1}+D(\ft92,\ft12)_1
+D(\ft{11}2,\ft12)_{-1}+D(\ft{11}2,\ft12)_1\nonumber\\
&&+D(4,0)_0+D(5,0)_0+D(5,0)_{-2}+D(5,0)_2+D(6,0)_0.
\label{eq:n=2}
\end{eqnarray}
Replacing $S^7$ by $SE_7$ amounts to replacing $CP^3$ by an appropriate
Kahler-Einstein base $B$.  In this case, the internal isometry is generically
reduced from $\mathrm{SU}(4)\times\mathrm{U}(1)$.  Nevertheless, the notion of
truncating to
SU(4) singlets may simply be replaced by the prescription of truncating to
zero modes on the base $B$.  This procedure was in fact done in
\cite{Gauntlett:2009zw}, which constructed the non-linear Kaluza-Klein
reduction for all the bosonic fields contained in the above supermultiplets
(\ref{eq:n=0}) and (\ref{eq:n=2}) and furthermore verified the $\mathcal N=2$
supersymmetry.

For the case of IIB supergravity compactified on $SE_5$, it is straightforward
to generalize the squashed $S^5$ conjecture of \cite{Liu:2000gk}.  In this
case, however, the Kaluza-Klein spectrum is more involved, and is given
in Table~\ref{tbl:se5}.  A curious feature shows up here in that an
additional LH+RH chiral matter multiplet shows up at the `massless'
Kaluza-Klein level.  The $E_0=4$ scalar in this multiplet corresponds to
the IIB axi-dilaton, while the additional $E_0=3$ charged scalar is
precisely the charged scalar constructed in the holographic model of
\cite{Gubser:2009qm}.  At the higher Kaluza-Klein levels, the breathing
and squashing mode scalars correspond to the $E_0=8$ and $E_0=6$ scalars
in the massive vector multiplet.  In addition, consistent truncations
involving the $E_0=5$ ($m^2L^2=8$) doublet of vectors in the semi-long LH+RH
massive gravitino multiplet and the $E_0=7$ ($m^2L^2=24$) vector in the
massive vector multiplet were constructed in
\cite{Herzog:2008wg,Maldacena:2008wh,Adams:2008wt} in the context of
investigating non-relativistic conformal backgrounds in string theory.

\begin{table}
\begin{tabular}{l|l|l|l}
$n$&Multiplet&$SU(2,2|1)$&$SO(2,4)\times U(1)$\\
\hline
0&supergraviton&$\mathcal D(3,\fft12,\fft12)_0$&$D(4,1,1)_0
+D(3\fft12,1,\fft12)_{-1}+D(3\fft12,\fft12,1)_1+D(3,\fft12,\fft12)_0$\\
0&LH chiral&$\mathcal D(3,0,0)_2$&$D(3\fft12,\fft12,0)_1+D(3,0,0)_2
+D(4,0,0)_0$\\
0&RH chiral&$\mathcal D(3,0,0)_{-2}$&$D(3\fft12,0,\fft12)_{-1}
+D(3,0,0)_{-2}+D(4,0,0)_0$\\
1&LH massive gravitino&$\mathcal D(4\fft12,0,\fft12)_1$&
$D(5\fft12,\fft12,1)_1+D(5,\fft12,\fft12)_0+D(5,0,1)_2$\\
&&&$\qquad+D(6,0,1)_0+D(4\fft12,0,\fft12)_1+D(5\fft12,0,\fft12)_{-1}$\\
1&RH massive gravitino&$\mathcal D(4\fft12,\fft12,0)_{-1}$&
$D(5\fft12,1,\fft12)_{-1}+D(5,\fft12,\fft12)_0+D(5,1,0)_{-2}$\\
&&&$\qquad+D(6,1,0)_0+D(4\fft12,\fft12,0)_{-1}+D(5\fft12,\fft12,0)_1$\\
2&massive vector&$\mathcal D(6,0,0)_0$&$D(7,\ft12,\ft12)_0
+D(6\ft12,\ft12,0)_{-1}+D(6\ft12,0,\ft12)_1$\\
&&&$\qquad+D(7\ft12,0,\ft12)_{-1}+D(7\ft12,\ft12,0)_1+D(6,0,0)_0$\\
&&&$\qquad+D(7,0,0)_{-2}+D(7,0,0)_2+D(8,0,0)_0$
\end{tabular}
\caption{The truncated Kaluza-Klein spectrum of IIB supergravity on
squashed $S^5$ \cite{Liu:2000gk}, or equivalently on $SE_5$.
Here $n$ denotes the Kaluza-Klein level.  The consistent truncation is
expected to terminate at level $n=2$ with the breathing mode supermultiplet.}
\label{tbl:se5}
\end{table}

What we have seen so far is that massive consistent truncations of IIB
supergravity have been obtained keeping various subsets of the bosonic
fields identified in Table~\ref{tbl:se5}.  The goal of this paper is to
construct a complete non-linear Kaluza-Klein reduction of IIB supergravity
on $SE_5$ retaining all the bosonic fields in the multiplets up to the $n=2$
level.  This complements the massive Kaluza-Klein truncation of $D=11$
supergravity \cite{Gauntlett:2009zw}, and provides another example of a
consistent truncation retaining the breathing mode supermultiplet.
We proceed in Section~\ref{sec:iib} with the Sasaki-Einstein reduction of
IIB supergravity.  Then in Section~\ref{sec:linkk} we connect the full
non-linear reduction with the linearized Kaluza-Klein analysis of
\cite{Gunaydin:1984fk,Kim:1985ez} and show how the bosonic fields in
Table~\ref{tbl:se5} are related to the original IIB fields.  In
Section~\ref{sec:further} we relate the complete non-linear reduction to
previous results by performing additional truncations to a subset of active
fields.  Finally, we conclude in Section~\ref{sec:discuss} with some
further speculation on massive consistent truncations of supergravity.

While this work was being completed we became aware of
\cite{Cassani:2010uw,Gauntlett:2010vu,Skenderis:2010vz}
which independently worked out the massive consistent truncation of IIB
supergravity on $SE_5$.  

\section{Sasaki-Einstein reduction of IIB supergravity}
\label{sec:iib}

The bosonic field content of IIB supergravity consists of the NSNS fields
$(g_{MN},B_{MN},\phi)$ and the RR potentials $(C_0,C_2,C_4)$.  Because of
the self-dual field strength $F_5^+=dC_4$, it is not possible to write down
a covariant action.  However, we may take a bosonic Lagrangian of the form
\begin{equation}
\mathcal L_{\mathrm{IIB}}=R*1-\fft1{2\tau_2^2}d\tau\wedge*d\bar\tau
-\fft12\mathcal M_{ij}F_3^i
\wedge*F_3^j-\fft14\widetilde F_5\wedge*\widetilde F_5-\fft14\epsilon_{ij}
C_4\wedge F_3^i\wedge F_3^j,
\label{eq:iiblag}
\end{equation}
where self-duality $\widetilde F_5=*\widetilde F_5$ is to be imposed by
hand after deriving the equations of motion.

We have given the Lagrangian in an SL(2,$\mathcal R$) invariant form where
\begin{equation}
\tau=C_0+ie^{-\phi},\qquad
\mathcal M=\fft1{\tau_2}\begin{pmatrix}|\tau|^2&-\tau_1\cr-\tau_1&1
\end{pmatrix},
\end{equation}
and where
\begin{equation}
F_3^i=dB_2^i,\qquad B_2^i=\begin{pmatrix}B_2\cr C_2\end{pmatrix},\qquad
\widetilde F_5=dC_4+\ft12\epsilon_{ij}B_2^i\wedge dB_2^j.
\end{equation}
The equations of motion following from (\ref{eq:iiblag}) and the self-duality
of $\widetilde F_5$ are
\begin{eqnarray}
d\widetilde F_5&=&\ft12\epsilon_{ij}F_3^i\wedge F_3^j,\qquad
\widetilde F_5=*\widetilde F_5,\nonumber\\
d(\mathcal M_{ij}*F_3^j)&=&-\epsilon_{ij}\widetilde F_5\wedge F_3^j,
\nonumber\\
\fft{d*d\tau}{\tau_2}+i\fft{d\tau\wedge*d\tau}{\tau_2^2}
&=&-\fft{i}{2\tau_2}G_3\wedge*G_3,
\label{eq:taueom}
\end{eqnarray}
and the Einstein equation (in Ricci form)
\begin{eqnarray}
R_{MN}&=&\fft1{2\tau_2^2}\partial_{(M}\tau\partial_{N)}\bar\tau
+\fft14\mathcal M_{ij}\left(F_{MPQ}^iF_N^{j\,PQ}-\fft1{12}g_{MN}F_{PQR}^i
F^{j\,PQR}\right)\nonumber\\
&&+\fft1{4\cdot4!}\widetilde F_{MPQRS}\widetilde F_N{}^{PQRS}.
\label{eq:10eins}
\end{eqnarray}
In the above we have introduced the complex three-form $G_3=F^2_3-\tau F^1_3$.
If desired, this allows us to rewrite the three-form equation of motion as
\begin{equation}
d*G=-i\fft{d\tau}{2\tau_2}\wedge*(G_3+\bar G_3)+i\widetilde F_5\wedge G_3.
\end{equation}

\subsection{The reduction ansatz}

Before writing out the reduction ansatz, we note a few key features of
Sasaki-Einstein manifolds. A Sasaki-Einstein manifold has a preferred U(1)
isometry related to the Reeb vector.  This allows us to write the metric as
a U(1) fibration over a Kahler-Einstein base $B$
\begin{equation}
ds^2(SE_5)=ds^2(B)+(d\psi+\mathcal A)^2,
\end{equation}
where $d\mathcal A=2J$ with $J$ the Kahler form on $B$.  Moreover, $B$
admits an SU(2) structure defined by the (1,1) and (2,0) forms $J$ and $\Omega$
satisfying
\begin{equation}
J\wedge\Omega=0,\qquad \Omega\wedge\bar\Omega=2J\wedge J=4*_41,\qquad
*_4J=J,\qquad*_4\Omega=\Omega,
\end{equation}
as well as
\begin{equation}
dJ=0,\qquad d\Omega=3i(d\psi+\mathcal A)\wedge\Omega.
\end{equation}
Note that we are taking the `unit radius' Einstein condition
$R_{ij}=4g_{ij}$ on the Sasaki-Einstein manifold, which corresponds to
$R_{ab}=6g_{ab}$ on the Kahler-Einstein base.


For the reduction, we write down the most general decomposition of the
bosonic IIB fields consistent with the isometries of $B$.  For the metric,
we take
\begin{equation}
ds_{10}^2=e^{2A}ds_5^2+e^{2B}ds^2(B)+e^{2C}(\eta+A_1)^2,
\label{eq:metans}
\end{equation}
where $\eta=d\psi+\mathcal A$.  Since $A_1$ gauges the U(1) isometry, it
will be related to the $D=5$ graviphoton.  Note, however, that the graviphoton
receives additional contributions from the five-form.

The three-form and five-form field strengths can be expanded in a basis of
invariant tensors on $B$.  For the three-forms, we work with the potentials
\begin{equation}
B_2^i=b^i_2+b^i_1\wedge(\eta+A_1)+b^i_0\Omega+\bar b^i_0\bar\Omega.
\label{eq:b2ans}
\end{equation}
The scalars $b^i_0$ are complex, while the remaining fields are real. Note that we do not include a term of the form $\widetilde b^i_0 J$ in the ansatz, as this field will act simply as a St$\ddot{u}$ckelburg field in the five-dimensional theory. In particular, it does not give rise to any new dynamics in the equations of motion as it can be repackaged as a total derivative plus terms which would simply shift $b^i_2$ and $b^i_1$,
\begin{equation}
2\widetilde b^i_0 J  = d(\widetilde b^i_0\wedge(\eta+A_1)) - d\widetilde b^i_0\wedge(\eta+A_1) - \widetilde b^i_0 F_2.
\end{equation}
Taking $F^i_3=dB^i_2$ gives
\begin{eqnarray}
F^i_3&=&(db^i_2-b^i_1\wedge F)+db^i_1\wedge(\eta+A_1)-2b^i_1\wedge J
+Db^i_0\wedge\Omega+D\bar b^i_0\wedge\bar\Omega\nonumber\\
&&+3ib^i_0\Omega\wedge(\eta+A_1)-3i\bar b^i_0\bar\Omega\wedge(\eta+A_1),
\label{eq:f3ans0}
\end{eqnarray}
where $D$ is the U(1) gauge covariant derivative
\begin{equation}
Db_0^i=db_0^i-3iA_1b_0^i.
\end{equation}
For convenience, we write this as
\begin{equation}
F_3^i=g_3^i+g_2^i\wedge(\eta+A_1)+g_1^i\wedge J+f_1^i\wedge\Omega
+\bar f_1^i\wedge\bar\Omega+f_0^i\wedge\Omega\wedge(\eta+A_1)
+\bar f_0^i\wedge\bar\Omega\wedge(\eta+A_1),
\label{eq:f3ans}
\end{equation}
where our notation is such that the $g^i$'s are real and the $f^i$'s are
complex.

For the self-dual five-form, we take
\begin{equation}
\widetilde F_5=(1+*)[(4+\phi_0)*_41\wedge(\eta+A_1)
+\mathbb A_1\wedge*_41+p_2\wedge J\wedge(\eta+A_1)
+q_2\wedge\Omega\wedge(\eta+A_1)+\bar q_2\wedge\bar\Omega\wedge(\eta+A_1)],
\end{equation}
where $*_41$ denotes the volume form on the Kahler-Einstein base $B$.
Note that we have pulled out a constant background component
\begin{equation}
\widetilde F_5=4(1+*)\mathrm{vol}(SE_5),
\end{equation}
which sets up the Freund-Rubin compactification%
\footnote{For simplicity, we have assumed a unit radius ($L=1$)
compactification.}.
The two-forms $q_2$
are complex, while the other fields are real.  For later convenience,
we take the explicit 10-dimensional dual in the metric (\ref{eq:metans})
to obtain
\begin{eqnarray}
\widetilde F_5&=&(4+\phi_0)*_41\wedge(\eta+A_1)
+\mathbb A_1\wedge*_41+p_2\wedge J\wedge(\eta+A_1)
+q_2\wedge\Omega\wedge(\eta+A_1)\nonumber\\
&&+\bar q_2\wedge\bar\Omega\wedge(\eta+A_1)
+e^{5A-4B-C}(4+\phi_0)*1-e^{3A-4B+C}*\mathbb A_1\wedge(\eta+A_1)\nonumber\\
&&+e^{A-C}*p_2\wedge J
+e^{A-C}*q_2\wedge\Omega+e^{A-C}*\bar q_2\wedge\bar\Omega,
\label{eq:f5ans}
\end{eqnarray}
where $*$ now denotes the Hodge dual in the $D=5$ spacetime.

\subsection{Reduction of the equations of motion}

In order to obtain the reduction, it is now simply a matter of inserting
the above decompositions into the IIB equations of motion.  The $\widetilde F_5$
equation yields
\begin{eqnarray}
d(e^{A-C}*p_2)&=&2e^{3A-4B+C}*\mathbb A_1-p_2\wedge F_2+\epsilon_{ij}g_1^i\wedge
g_3^j,\nonumber\\
Dq_2&=&3ie^{A-C}*q_2+\epsilon_{ij}(f_1^i\wedge g_2^j-f_0^ig_3^j),
\label{eq:f5eom}
\end{eqnarray}
along with the constraints
\begin{eqnarray}
\phi_0&=&-\ft{2i}3\epsilon_{ij}(f_0^i\bar f_0^j-\bar f_0^if_0^j),\nonumber\\
p_2&=&\ft14\epsilon_{ij}g_1^i\wedge g_1^j-d[A_1+\ft14\mathbb A_1+\ft{i}6
\epsilon_{ij}(f_0^i\bar f_1^j-\bar f_0^if_1^j)].
\label{eq:cons}
\end{eqnarray}
The implication of this is that $\widetilde F_5$ gives rise to two physical
$D=5$ fields, namely a massive vector $\mathbb A_1$ and a
complex antisymmetric tensor $q_2$ satisfying an odd-dimensional self-duality
equation and with $m^2=9$.  The mass of $\mathbb A_1$ is not directly
apparent from (\ref{eq:f5eom}) as it mixes with $A_1$ from the metric to
yield the massless graviphoton as well as a $m^2=24$ massive vector.

The $F_3^i$ equation yields
\begin{eqnarray}
D(e^{3A+C}\mathcal M_{ij}*f_1^j)\!\!&=&\!\!-3ie^{5A-C}\mathcal M_{ij}f_0^j*1
+\epsilon_{ij}[(4+\phi_0)e^{5A-4B-C}f_0^j*1-q_2\wedge g_3^j\nonumber\\
&&\!\!+e^{A-C}*q_2\wedge g_2^j+e^{3A-4B+C}*\mathbb A_1\wedge f_1^j],\nonumber\\
d(e^{A+4B-C}\mathcal M_{ij}*g_2^j)\!\!&=&\!\!
\mathcal M_{ij}[e^{-A+4B+C}*g_3^j\wedge F+4e^{3A+C}*g_1^j]\nonumber\\
&&\!\!+\epsilon_{ij}[-2e^{A-C}*p_2\wedge g_1^j-\mathbb A_1\wedge g_3^j-4e^{A-C}
(*q_2\wedge\bar f_1^j+*\bar q_2\wedge f_1^j)],\nonumber\\
d(e^{-A+4B+C}\mathcal M_{ij}*g_3^j)\!\!&=&\!\!\epsilon_{ij}[-(4+\phi_0)g_3^j
+\mathbb A_1\wedge g_2^j-2p_2\wedge g_1^j-4(q_2\wedge\bar f_1^j
+\bar q_2\wedge f_1^j)\nonumber\\
&&\!\!+4e^{A-C}(\bar f_0^j*q_2+f_0^j*\bar q_2)].
\label{eq:f3eom}
\end{eqnarray}
These correspond to a pair of charged scalars $f_0^i$, a pair of $m^2=8$
massive vectors $g_1^i$ and a pair of massive antisymmetric tensors $b_2^i$.

The ten-dimensional Einstein equation (\ref{eq:10eins}) reduces to a
five-dimensional Einstein equation, as well as the equations of motion
for the breathing and squashing modes $B$ and $C$ and the graviphoton
$A_1$.  In particular, in the natural vielbein basis, the frame components
of the ten-dimensional
Ricci tensor corresponding to the reduction (\ref{eq:metans}) are given by
\begin{eqnarray}
{}^{10}R_{\alpha\beta}&=&e^{-2A}[R_{\alpha\beta}-\nabla_\alpha\nabla_\beta
(3A+4B+C)-\eta_{\alpha\beta}\partial_\gamma A\partial^\gamma(3A+4B+C)
-\eta_{\alpha\beta}\square A\nonumber\\
&&+3\partial_\alpha A\partial_\beta A
-4\partial_\alpha B\partial_\beta B-\partial_\alpha C\partial_\beta C
+4(\partial_\alpha A\partial_\beta B+\partial_\alpha B\partial_\beta A)
\nonumber\\
&&+(\partial_\alpha A\partial_\beta C+\partial_\alpha C\partial_\beta A)]
-\ft12e^{2C-4A}F_{\alpha\gamma}F_\beta{}^\gamma,\nonumber\\
{}^{10}R_{ab}&=&\delta_{ab}[6e^{-2B}-2e^{2C-4B}-e^{-2A}(\square B
+\partial_\gamma B\partial^\gamma(3A+4B+C))],\nonumber\\
{}^{10}R_{99}&=&4e^{2C-4B}+\ft14e^{2C-4A}F_{\gamma\delta}F^{\gamma\delta}
-e^{-2A}(\square C+\partial_\gamma C\partial^\gamma(3A+4B+C)),\nonumber\\
{}^{10}R_{\alpha9}&=&\ft12e^{C-3A}[\nabla^\gamma F_{\alpha\gamma}
+F_{\alpha\gamma}\partial^\gamma(A+4B+3C)].
\label{eq:10ricci}
\end{eqnarray}
The $\alpha$ and $\beta$ indices correspond to the $D=5$ spacetime, while
$a$ and $b$ correspond to the Kahler-Einstein base $B$ and $9$ corresponds to
the U(1) fiber direction.  The covariant derivatives and frame indices on
the right hand side of these quantities are with respect to the $D=5$ metric.
In order to reduce to the $D=5$ Einstein frame metric, we now choose
$3A+4B+C=0$, or
\begin{equation}
A=-\ft43B-\ft13C.
\label{eq:aconst}
\end{equation}
For convenience, we will retain $A$ in the expressions below.  However, it
is not independent, and should always be thought of as a shorthand for
(\ref{eq:aconst}).

Equating the ten-dimensional Ricci tensor (\ref{eq:10ricci}) to the stress
tensor formed out of $F_3^i$ and $\widetilde F_5$ of (\ref{eq:f3ans}) and
(\ref{eq:f5ans}), we obtain the $D=5$ Einstein equation
\begin{eqnarray}
R_{\alpha\beta}&=&
\ft13\eta_{\alpha\beta}(-24e^{2A-2B}+4e^{5A+3C}+\ft12e^{8A}
(4+\phi_0)^2)+\ft{28}3\partial_\alpha B\partial_\beta B
+\ft83\partial_{(\alpha}B\partial_{\beta)}C\nonumber\\
&&+\ft43\partial_\alpha C\partial_\beta C+\ft1{2\tau_2^2}\partial_{(\alpha}\tau
\partial_{\beta)}\bar\tau
+\ft12e^{2C-2A}(F_{\alpha\gamma}F_\beta{}^\gamma
-\ft16\eta_{\alpha\beta}F_{\gamma\delta}F^{\gamma\delta})
+\ft12e^{-8B}\mathbb A_\alpha\mathbb A_\beta\nonumber\\
&&+e^{A-C}[(p_{\alpha\gamma}p_\beta{}^\gamma
-\ft16\eta_{\alpha\beta}p_{\gamma\delta}p^{\gamma\delta})
+4(q_{(\alpha}{}^\gamma\bar q_{\beta)\gamma}-\ft16\eta_{\alpha\beta}
q_{\gamma\delta}\bar q^{\gamma\delta})]\nonumber\\
&&+\mathcal M_{ij}[\ft23e^{5A-C}\eta_{\alpha\beta}(f_0^i\bar f_0^j
+\bar f_0^if_0^j)+\ft12e^{-2A-2C}(g_{\alpha\gamma}^ig_\beta^{j\,\gamma}
-\ft16\eta_{\alpha\beta}g_{\gamma\delta}^ig^{j\,\gamma\delta})\nonumber\\
&&+\ft14e^{-4A}(g_{\alpha\gamma\delta}^ig_\beta^{j\,\gamma\delta}
-\ft29\eta_{\alpha\beta}g_{\gamma\delta\epsilon}^ig^{j\,\gamma\delta\epsilon})
+e^{-4B}(g_\alpha^ig_\beta^j+2(f_\alpha^i\bar f_\beta^j
+\bar f_\alpha^if_\beta^j))],
\label{eq:einseom}
\end{eqnarray}
as well as the $B$, $C$ and $A_1$ equations of motion
\begin{eqnarray}
d*dB&=&[6e^{2A-2B}-2e^{5A+3C}-\ft14e^{8A}(4+\phi_0)^2]*1-\ft14e^{-8B}
\mathbb A_1\wedge*\mathbb A_1\nonumber\\
&&+\mathcal M_{ij}[\ft18e^{-2A-2C}g_2^i\wedge*g_2^j
+\ft18e^{-4A}g_3^i\wedge*g_3^j-\ft12e^{5A-C}(f_0^i\bar f_0^j
+\bar f_0^if_0^j)*1\nonumber\\
&&-\ft14e^{-4B}(g_1^i\wedge*g_1^j+2(f_1^i\wedge*\bar f_1^j
+\bar f_1^i\wedge*f_1^j))],\nonumber\\
d*dC&=&[4e^{5A+3C}-\ft14e^{8A}(4+\phi_0)^2]*1
+\ft12e^{2C-2A}F_2\wedge*F_2+\ft14e^{-8B}\mathbb A_1\wedge*\mathbb A_1
\nonumber\\
&&-\ft12e^{A-C}(p_2\wedge*p_2+4q_2\wedge*\bar q_2)
+\mathcal M_{ij}[-\ft38e^{-2A-2C}g_2^i\wedge*g_2^j\nonumber\\
&&+\ft18e^{-4A}g_3^i\wedge*g_3^j-\ft32e^{5A-C}(f_0^i\bar f_0^j
+\bar f_0^if_0^j)*1\nonumber\\
&&+\ft14e^{-4B}(g_1^i\wedge*g_1^j+2(f_1^i\wedge*\bar f_1^j
+\bar f_1^i\wedge*f_1^j))],\nonumber\\
d(e^{2C-2A}*F_2)&=&(4+\phi_0)e^{-8B}*\mathbb A_1-p_2\wedge p_2
-4q_2\wedge\bar q_2\nonumber\\
&&+\mathcal M_{ij}[4e^{-4B}*(f_0^i\bar f_1^j+\bar f_0^if_1^j)
+e^{-4A}*g_3^i\wedge g_2^j].
\label{eq:eins2eom}
\end{eqnarray}
Note that, in order to obtain the $D=5$ Einstein equation, we had to
shift the reduction of ${}^{10}R_{\alpha\beta}$ an appropriate combination
of ${}^{10}R_{ab}$ and ${}^{10}R_{99}$ in order to remove the
$\eta_{\alpha\beta}\square A$ component in the first line of
(\ref{eq:10ricci}).

The IIB equations of motion thus reduce to (\ref{eq:f5eom}), (\ref{eq:f3eom}),
(\ref{eq:einseom}) and (\ref{eq:eins2eom}) as well as the axi-dilaton equation,
which we have not written down explicitly, but which will be shown to be
consistent below.

\subsection{The effective five-dimensional Lagrangian}

We now wish to construct an effective $D=5$ Lagrangian which reproduces the above
equations of motion.  This may be done by noting that the $D=5$ Einstein
equation (\ref{eq:einseom}) arises naturally from a Lagrangian of the form
\begin{eqnarray}
\mathcal L&=&R*1+(24e^{2A-2B}-4e^{5A+3C}-\ft12e^{8A}(4+\phi_0)^2)*1
-\ft{28}3dB\wedge*dB-\ft83dB\wedge*dC\nonumber\\
&&-\ft43dC\wedge*dC-\ft1{2\tau_2^2}d\tau\wedge*d\bar\tau
-\ft12e^{2C-2A}F_2\wedge*F_2-\ft12e^{-8B}\mathbb A_1\wedge*\mathbb A_1
\nonumber\\
&&-e^{A-C}(p_2\wedge*p_2+4q_2\wedge*\bar q_2)+\mathcal M_{ij}
[-2e^{5A-C}(f_0^i\bar f_0^j+\bar f_0^if_0^j)*1\nonumber\\
&&-\ft12e^{-2A-2C}g_2^i\wedge*g_2^j-\ft12e^{-4A}g_3^i\wedge*g_3^j
-e^{-4B}(g_1^i\wedge*g_1^j+2(f_1^i\wedge*\bar f_1^j
+\bar f_1^i\wedge*f_1^j))]\nonumber\\
&&+\mathcal L_{CS}.
\label{eq:lag0}
\end{eqnarray}
We have included a Chern-Simons piece $\mathcal L_{CS}$ which cannot be
determined from the Einstein equation.

It is now possible to verify that (\ref{eq:lag0}) reproduces all the terms
in the equations of motion (\ref{eq:f5eom}), (\ref{eq:f3eom}) and
(\ref{eq:eins2eom}) involving the metric ({\it ie} the Hodge *).  The
remaining terms may be obtained from the addition of the topological piece
\begin{eqnarray}
\mathcal L_{CS}&=&\ft{2i}3(q_2\wedge d\bar q_2-\bar q_2\wedge dq_2)
-4A_1\wedge q_2\wedge\bar q_2+2\epsilon_{ij}b_2^i\wedge db_2^j\nonumber\\
&&+\ft{4i}3[(\bar q_2-\ft{i}6\epsilon_{ij}\bar f_0^ig_2^j)
\wedge\epsilon_{kl}(f_1^k\wedge g_2^l-f_0^kg_3^l)
-(q_2+\ft{i}6\epsilon_{ij}f_0^ig_2^j)
\wedge\epsilon_{kl}(\bar f_1^k\wedge g_2^l-\bar f_0^kg_3^l)]\nonumber\\
&&-A_1\wedge(p_2-\ft14\epsilon_{ij}g_1^i\wedge g_1^j)\wedge
(p_2-\ft14\epsilon_{kl}g_1^k\wedge g_1^l)\nonumber\\
&&-2[\ft14\mathbb A_1+\ft{i}6\epsilon_{ij}(f_0^i\bar f_1^j-\bar f_0^if_1^j)]
\wedge\epsilon_{kl}(g_1^k\wedge g_3^l-\ft14g_1^k\wedge g_1^l\wedge F_2).
\label{eq:lag1}
\end{eqnarray}
Here we recall the definitions
\begin{equation}
f_0^i=3ib_0^i,\qquad f_1^i=Db_0^i,\qquad
g_1^i=-2b_1^i,\qquad g_2^i=db_1^i,\qquad g_3^i=db_2^i-b_1^i\wedge F_2,
\end{equation}
implicit in (\ref{eq:f3ans0}) and (\ref{eq:f3ans}).  Furthermore, $\phi_0$
and $p_2$ are given by (\ref{eq:cons}).  Note that, while $\mathbb A_1$ is
massive, and does not have a gauge invariance associated with it, it is
natural to make the shift
\begin{equation}
\mathbb A_1\to\mathbb A'_1-\ft{2i}3\epsilon_{ij}(f_0^i\bar f_1^j
-\bar f_0^if_1^j),
\end{equation}
so that
\begin{equation}
p_2=\ft14\epsilon_{ij}g_1^i\wedge g_1^j-F_2-\ft14\mathbb F'_2,
\end{equation}
where $\mathbb F'_2=d\mathbb A'_1$.

We now turn to the axi-dilaton equation obtained from (\ref{eq:lag0}).
Since $\tau$ only shows up in the kinetic term and in $\mathcal M_{ij}$,
we see that the $\tau$ equation of motion obtained from the $D=5$ Lagrangian
reproduces that obtained from the original IIB Lagrangian.  This is because
the quantity in the square brackets multiplying $\mathcal M_{ij}$ in
(\ref{eq:lag0}) is the straightforward reduction of $-\fft12F_3^i\wedge*F_3^j$
in the original IIB Lagrangian (\ref{eq:iiblag}).

\section{Matching the linearized Kaluza-Klein analysis}
\label{sec:linkk}

The complete $D=5$ Lagrangian, as given by (\ref{eq:lag0}) and (\ref{eq:lag1}),
is somewhat opaque.  Thus in this section, we demonstrate that it in fact
contains the fields corresponding to the Kaluza-Klein mass spectrum noted
in Table~\ref{tbl:se5}.  To do this, it is sufficient to look at the linearized
level.  We first note that the effective $D=5$ fields are the complex scalars
$(\tau, b_0^i)$, real scalars $(B,C)$, one-form potentials
$(A_1,b_1^i,\mathbb A_1)$, pair of real two-forms $(b_2^i)$, the complex
two-form $(q_2)$, and of course the metric $(g_{\mu\nu})$.
The $D=5$ equations of motion (\ref{eq:f5eom}), (\ref{eq:f3eom}) and
(\ref{eq:eins2eom}) may be linearized on the matter fields to obtain the set
\begin{eqnarray}
d*db_0^i&=&(9\delta^i_j+12i\mathcal N^i{}_j)b_0^j*1,\nonumber\\
d*db_1^i&=&-8*b_1^i,\nonumber\\
d*db_2^i&=&-4\mathcal N^i{}_jdb_2^j,\nonumber\\
dq_2&=&3i*q_2,\nonumber\\
d*F_2&=&4*\mathbb A_1,\kern4em d*F_2+\ft14d*\mathbb F_2=-2*\mathbb A_1,
\nonumber\\
d*dB&=&4(7B+C)*1,\qquad d*dC=16(B+C)*1.
\label{eq:lineom}
\end{eqnarray}
Here we have introduced
\begin{equation}
\mathcal N=\mathcal M^{-1}\epsilon=\fft1{\tau_2}\begin{pmatrix}-\tau_1&1\cr
-|\tau|^2&\tau_1\end{pmatrix},
\end{equation}
with eigenvalues $+i$ and $-i$, corresponding to eigenvectors
$\begin{pmatrix}1&\tau\end{pmatrix}^T$ and
$\begin{pmatrix}1&\bar\tau\end{pmatrix}^T$, respectively.

The first equation in (\ref{eq:lineom}) then decomposes into a pair of
equations for the complex scalars $b_0^{m^2=-3}$ and $b_0^{m^2=21}$
with masses $m^2=-3$ and $m^2=21$ according to
\begin{equation}
b_0^i=\begin{pmatrix}1\cr\tau\end{pmatrix}b_0^{m^2=-3}
+\begin{pmatrix}1\cr\bar\tau\end{pmatrix}b_0^{m^2=21}.
\end{equation}
The second equation is that of an SL(2,$\mathbb R$) doublet of real
vectors $b_1^i$ with mass $m^2=8$.  The third equation can be converted to
an odd-dimensional self-duality equation \cite{Townsend:1983xs}
$db_2^i=4\mathcal N^i{}_j*b_2^j$,
for a doublet of antisymmetric tensors $b_2^i$ with mass $m^2=16$.  The
fourth equation is already in odd-dimensional self-duality form, and
shows that the complex antisymmetric tensor $q_2$ has mass $m^2=9$.

The vector equations can be diagonalized
\begin{equation}
d*(F_2+\ft16\mathbb F_2)=0,\qquad d*\mathbb F_2=-24*\mathbb A_1,
\end{equation}
to identify the massless graviphoton $A_1+\fft16\mathbb A_1$ and the
massive $m^2=24$ vector $\mathbb A_1$.  Finally the $B$ and $C$ equations
may be diagonalized to identify the $m^2=32$ breathing and $m^2=12$
squashing modes
\begin{equation}
d*d\rho=32\rho*1,\qquad d*d\sigma=12\sigma*1,
\end{equation}
where
\begin{equation}
B=\rho+\ft12\sigma,\qquad C=\rho-2\sigma.
\end{equation}

It is now possible to see how the above linearized modes are organized into
$\mathcal N=2$ supermultiplets.  As shown in Table~\ref{tbl:se5}, at the zeroth
Kaluza-Klein level, we have the graviton supermultiplet
\begin{equation}
\mathcal D(3,\ft12,\ft12)_0=D(4,1,1)_0+D(3\ft12,1,\ft12)_{-1}
+D(3\ft12,\ft12,1)_1+D(3,\ft12,\ft12)_0,
\end{equation}
with bosonic fields being the graviton $g_{\mu\nu}$ and the massless
graviphoton $A_1+\fft16\mathbb A_1$.  Still at the zeroth level, there
is also a LH+RH chiral multiplet
\begin{eqnarray}
\mathcal D(3,0,0)_2&=&D(3\ft12,\ft12,0)_1+D(3,0,0)_2+D(4,0,0)_0,\nonumber\\
\mathcal D(3,0,0)_{-2}&=&D(3\ft12,0,\ft12)_{-1}+D(3,0,0)_{-2}+D(4,0,0)_0.
\end{eqnarray}
The charged $E_0=3$ scalar corresponds to the $m^2=-3$ scalar $b_0^{m^2=-3}$,
while the complex $E_0=4$ scalar is the axi-dilaton $\tau$.

At the first Kaluza-Klein level, we have a semi-long LH+RH massive gravitino
multiplet
\begin{eqnarray}
\mathcal D(4\ft12,0,\ft12)_1&=&D(5\ft12,\ft12,1)_1+D(5,\ft12,\ft12)_0
+D(5,0,1)_2+D(6,0,1)_0\nonumber\\
&&\qquad+D(4\ft12,0,\ft12)_1+D(5\ft12,0,\ft12)_{-1},\nonumber\\
\mathcal D(4\ft12,\ft12,0)_{-1}&=&D(5\ft12,1,\ft12)_{-1}+D(5,\ft12,\ft12)_0
+D(5,1,0)_{-2}+D(6,1,0)_0\nonumber\\
&&\qquad+D(4\ft12,\ft12,0)_{-1}+D(5\ft12,\ft12,0)_1.
\end{eqnarray}
The bosonic field content is an SL(2,$\mathbb R$) doublet of $m^2=8$
($E_0=5$) vectors $b_1^i$, a charged $m^2=9$ ($E_0=5$) anti-symmetric
tensor $q_2$, and a doublet of $m^2=16$ ($E_0=6$) anti-symmetric
tensors $b_2^i$.

At the second Kaluza-Klein level, we have a massive vector multiplet
\begin{eqnarray}
\mathcal D(6,0,0)_0&=&D(7,\ft12,\ft12)_0+D(6\ft12,\ft12,0)_{-1}
+D(6\ft12,0,\ft12)_1+D(7\ft12,0,\ft12)_{-1}+D(7\ft12,\ft12,0)_1\nonumber\\
&&\qquad+D(6,0,0)_0+D(7,0,0)_{-2}+D(7,0,0)_2+D(8,0,0)_0.
\end{eqnarray}
The massive $E_0=7$ vector is the $m^2=24$ mode $\mathbb A_1$.
The real $E_0=6$ and $E_0=8$ scalars are the $m^2=12$ squashing and
$m^2=32$ breathing modes, $\sigma$ and $\rho$, respectively.  The charged
$E_0=7$ scalar is $b_0^{m^2=21}$ with $m^2=21$.  This identification of the
linearized fields with the Kaluza-Klein modes is shown in
Table~\ref{tbl:match}.

\begin{table}
\begin{tabular}{l|l|l|l}
n&Multiplet&State&Field\\
\hline
0&supergraviton&$D(4,1,1)_0$&$g_{\mu\nu}$\\
&&$D(3,\fft12,\fft12)_0$&$A_1+\fft16\mathbb A_1$\\
\hline
0&LH+RH chiral&$D(3,0,0)_\pm2$&$b_0^{m^2=-3}$\\
&&$D(4,0,0)_0+D(4,0,0)_0$&$\tau$\\
\hline
1&LH+RH massive gravitino&$D(5,\fft12,\fft12)_0+D(5,\fft12,\fft12)_0$&$b_1^i$\\
&&$D(5,0,1)_2+D(5,1,0)_{-2}$&$q_2$\\
&&$D(6,0,1)_0+D(6,1,0)_0$&$b_2^i$\\
\hline
2&massive vector&$D(7,\fft12,\fft12)_0$&$\mathbb A_1$\\
&&$D(6,0,0)_0$&$\sigma$\\
&&$D(7,0,0)_\pm2$&$b_0^{m^2=21}$\\
&&$D(8,0,0)_0$&$\rho$\\
\end{tabular}
\caption{Identification of the bosonic states in the Kaluza-Klein spectrum
with the linearized modes in the reduction.}
\label{tbl:match}
\end{table}

For the case of IIB supergravity on $S^5$, is interesting to note that these
fields lie at the lowest level of the massive trajectories in the Kaluza-Klein
mode decomposition of the $D=10$ fields \cite{Gunaydin:1984fk,Kim:1985ez}.
We note that the massive Kaluza-Klein tower is built out of scalar, vector
and tensor harmonics on $S^5$, and the lowest harmonics generally have simple
behavior on the internal sphere coordinates.  For example, the lowest
scalar harmonic is the constant mode on the sphere, while the lowest vector
harmonics generate the Killing vectors on the sphere.  It is presumably the
simplicity of the lowest harmonics that allows the truncation to be consistent,
even at the non-linear level.

Although the harmonics on $SE_5$ are more involved (see {\it e.g.}
\cite{Ceresole:1999zs} for the case of $T^{1,1}$), it is clear that the
decomposition (\ref{eq:b2ans}) and (\ref{eq:f5ans}) of the $D=10$ fields
in terms of invariant tensors on $SE_5$ is equivalent to the truncation
to the lowest harmonics on the sphere.  This appears to be an essential
feature guaranteeing the consistency of the massive truncation, and hence
we do not expect to be able to keep any additional multiplets in the
Kaluza-Klein tower beyond the $n=2$ level.

\section{Further truncations}
\label{sec:further}

In order to make a connection with previous results on massive consistent
truncations of IIB supergravity, we note that the semi-long LH+RH massive
gravitino multiplet at the first Kaluza-Klein level may be truncated out
by setting
\begin{equation}
b_1^i=0,\qquad b_2^i=0,\qquad q_2=0.
\end{equation}
It is easy to see that this truncation is consistent, since the
respective equations of motion for $q_2$ in (\ref{eq:f5eom}) and
$g_2^i$ and $g_3^i$ in (\ref{eq:f3eom}) are trivially satisfied in this case.
The resulting $D=5$ Lagrangian takes the form
\begin{eqnarray}
\mathcal L&=&R*1+(24e^{2A-2B}-4e^{5A+3C}-\ft12e^{8A}(4+\phi_0)^2)*1
-\ft{28}3dB\wedge*dB-\ft83dB\wedge*dC\nonumber\\
&&-\ft43dC\wedge*dC-\ft1{2\tau_2^2}d\tau\wedge*d\bar\tau
-\ft12e^{2C-2A}F_2\wedge*F_2-e^{A-C}(F_2+\ft14\mathbb F'_2)\wedge*
(F_2+\ft14\mathbb F'_2)\nonumber\\
&&-\ft12e^{-8B}[\mathbb A'_1-\ft{2i}3\epsilon_{ij}(f_0^i\bar f_1^j
-\bar f_0^if_1^j)]\wedge*[\mathbb A'_1
-\ft{2i}3\epsilon_{ij}(f_0^i\bar f_1^j-\bar f_0^if_1^j)]\nonumber\\
&&-2\mathcal M_{ij}
[e^{5A-C}(f_0^i\bar f_0^j+\bar f_0^if_0^j)*1
+e^{-4B}(f_1^i\wedge*\bar f_1^j+\bar f_1^i\wedge*f_1^j)]\nonumber\\
&&-A_1\wedge (F_2+\ft14\mathbb F'_2)\wedge (F_2+\ft14\mathbb F'_2),
\label{eq:non=1}
\end{eqnarray}
where
\begin{equation}
f_0^i=3ib_0^i,\qquad f_1^i=Db_0^i,\qquad
\phi_0=-\ft{2i}3\epsilon_{ij}(f_0^i\bar f_0^j-\bar f_0^if_0^j).
\end{equation}

A further truncation to the massless $\mathcal N=2$ supergravity sector
may be obtained by setting
\begin{equation}
b_0^i=0,\qquad B=0,\qquad C=0,\qquad \mathbb A_1=0,
\end{equation}
along with taking a constant background for the axi-dilaton, $\tau=\tau_0$.
This leaves only $g_{\mu\nu}$ and $A_1$, and yields the standard
Lagrangian for the bosonic fields of minimal gauged supergravity
\begin{equation}
\mathcal L = R*1 + 12g^2*1 - \ft12 F_2\wedge*F_2
-\ft1{3\sqrt{3}}A_1\wedge F_2\wedge F_2,
\end{equation}
where we have rescaled the graviphoton, $A_1 \rightarrow \ft1{\sqrt{3}}A_1 $,
so that it has a canonical kinetic term, and where we have restored the
dimensionful gauged supergravity coupling $g$.

\subsection{Truncation to the zeroth Kaluza-Klein level}

Beyond the truncation to minimal supergravity discussed above,
the first nontrivial truncation involves keeping only the lowest Kaluza-Klein level fields $\{\tau, b_0^{m^2=-3}\}$ dynamical. In what follows we will denote $b_0^{m^2=-3}$ simply as $b$ so that $(b_0^1,b_0^2) = (b,\tau b)$. This truncation is not as simple as setting all other fields to zero, as the equations of motion demand certain constraints to be satisfied. For this case
we start with the Lagrangian (\ref{eq:non=1}), obtained by setting
$b_2^i = b_1^i = q_2 = 0$.  We then impose the constraints
\begin{equation}
b_0^{m^2=21}=0,\qquad e^{4B}=e^{-4C} = 1-4\tau_2 |b|^2,\qquad
\mathbb A_1 = -4i\tau_2(bD\bar b - \bar bDb)
+ 4 |b|^2 d \tau_1.
\end{equation}
These in turn imply that
\begin{equation}
\phi_0 = -24\tau_2|b|^2,\qquad p_2 = -d A_1.
\end{equation}

To guarantee consistency, we have to check four constraints from the equations of motion (the $B$, $C$, $f_0^i$, and combined Maxwell Equation). They are all verified to hold identically, and hence the truncation to the supergravity plus the LH+RH chiral multiplet is consistent. The Lagrangian is given by
%
\begin{eqnarray}
\mathcal L&=&R*1+\left[24(1 - 3\tau_2|b|^2)e^{-4B}-4e^{-8B}-\ft12e^{-8B}(4+\phi_0)^2\right]*1
-8 dB\wedge*dB \nonumber\\
&& -\ft32F_2\wedge*F_2-\ft12e^{-8B}\mathbb A_1\wedge*\mathbb A_1
- 8e^{-4B}\tau_2Db\wedge*D\bar b - 2ie^{-4B}(\bar b Db \wedge*d\bar \tau - bD\bar b\wedge*d\tau)
\nonumber \\
&&-\ft1{2\tau_2^2}(1+8e^{-4B}\tau_2|b|^2)d\tau\wedge*d\bar\tau - A_1\wedge F_2 \wedge F_2.
\end{eqnarray}
This expression can be simplified by defining $\lambda \equiv 4\tau_2|b|^2$, giving
\begin{eqnarray}
\mathcal L&=&R*1+\fft{6(2-3\lambda)}{(1-\lambda)^2}*1
-\fft{d\lambda\wedge*d\lambda}{2(1-\lambda)^2}  - \fft{(1+\lambda)d\tau\wedge*d\bar\tau}{2(1-\lambda)\tau_2^2} -\fft32F_2\wedge*F_2-\fft{\mathbb A_1\wedge*\mathbb A_1}{2(1-\lambda)^2} \nonumber\\
&& - \fft{8\tau_2 Db\wedge*D\bar b}{1-\lambda} - \fft{2i}{1-\lambda}(\bar b Db\wedge*d\bar\tau - b D\bar b\wedge*d\tau) -A_1\wedge F_2 \wedge F_2.
\end{eqnarray}

If we further truncate the model by setting $\tau = i e^{-\phi_0} = ig_s^{-1}$,
which is consistent with the equation of motion for $\tau$ given in
(\ref{eq:taueom}), this reproduces the model used in \cite{Gubser:2009qm}
to describe a holographic superconductor using a $m^2=-3$ and $q=2$ charged
scalar. If we denote $b = \sqrt{g_s}f e^{i \theta}$, the truncated Lagrangian reads
\begin{eqnarray}
\mathcal L &=& R*1   - \ft32 F_2\wedge*F_2 - A_1\wedge F_2\wedge F_2 \nonumber\\
&& + 12\fft{(1-6f^2)}{(1-4f^2)^2}*1 - 8\fft{df\wedge*df}{(1-4f^2)^2}
- 8f^2\fft{ (d\theta -3A_1)\wedge*(d\theta -3A_1)}{(1-4f^2)^2}.
\end{eqnarray}
A further redefinition $f=\fft12\tanh\fft\eta2$ then reproduces the Lagrangian
given in \cite{Gubser:2009qm}.

\subsection{Truncation to the second Kaluza-Klein level}

Starting with the Lagrangian (\ref{eq:non=1}) with $b_2^i=b_1^i=q_2=0$,
it is possible to retain the $b_0^{m^2=21}$ scalar by setting $b_0^{m^2=-3}=0$.
In this case, we first let $b_0^2 = \bar \tau b_0^1$ and define
$b_0^1 = \sqrt{g_s}h e^{i\xi}$, so that $(h,\xi)$ describe the $m^2=21$
scalar.  Again, the scalar equations of motion lead to constraints, and in
particular the first equation in (\ref{eq:f3eom}) yields the equation of
motion for $h$ and $\xi$ as well as
\begin{equation}
d(e^{3A+C}*d\tau) + i e^{3A+C}\ft1\tau_2 d\tau\wedge*d\tau = 0.
\end{equation}
This is simply the $\tau$ equation of motion without any sources, and the
simplest thing to do is to set $\tau$ to be constant, $\tau = ie^{-\phi_0}
= ig_s^{-1}$. The remaining field content is then
$\{g_{\mu\nu}, A_1, \rho, \sigma, b_0^{m^2=21}, \mathbb A_1\}$,
corresponding to the supergravity multiplet coupled to the massive vector
multiplet.  It is now straightforward to complete the truncation,
and the Lagrangian is given by
\begin{eqnarray}
\mathcal L&=&R*1+\bigl(24e^{-\fft{16}3\rho-\sigma}-4e^{-\fft{16}3\rho-6\sigma}
-8e^{-\fft{40}3\rho}(1+6h^2)^2\bigr)*1
-\ft{40}3d\rho\wedge*d\rho - 5 d\sigma\wedge*d\sigma \nn \\
&&-\ft12e^{\fft{16}3\rho-4\sigma}F_2\wedge*F_2 - e^{-\fft83\rho+2\sigma}
(F_2+\ft14\mathbb F'_2)\wedge*(F_2+\ft14\mathbb F'_2)\nonumber\\
&&- \ft12e^{-8\rho-4\sigma}(\mathbb A'_1 + 8h^2\Gamma)\wedge*
(\mathbb A'_1 + 8h^2\Gamma )
- A_1\wedge (F_2+\ft14\mathbb F'_2)\wedge (F_2+\ft14\mathbb F'_2)\nonumber\\
&&-8\bigl(e^{-4\rho-2\sigma}dh\wedge*dh + e^{-4\rho-2\sigma}h^2\Gamma\wedge*\Gamma + e^{-\ft{28}3\rho+2\sigma}h^2*1\bigr),
\end{eqnarray}
where we have defined $\Gamma = d\xi - 3A_1$.

We can further truncate this by removing the $m^2=21$ scalar ({\it i.e.}
by setting $h=\xi=0$), giving the Lagrangian
\begin{eqnarray}
\mathcal L&=&R*1+(24e^{-\fft{16}3\rho-\sigma}-4e^{-\fft{16}3\rho-6\sigma}
-8e^{-\fft{40}3\rho})*1
-\ft{40}3d\rho\wedge*d\rho - 5 d\sigma\wedge*d\sigma \nn \\
&&-\ft12e^{\fft{16}3\rho-4\sigma}F_2\wedge*F_2 - e^{-\fft83\rho+2\sigma}
(F_2+\ft14\mathbb F'_2)\wedge*(F_2+\ft14\mathbb F'_2)
- \ft12e^{-8\rho-4\sigma}\mathbb A'_1\wedge*\mathbb A'_1\nonumber\\
&&- A_1\wedge (F_2+\ft14\mathbb F'_2)\wedge (F_2+\ft14\mathbb F'_2),
\end{eqnarray}
which corresponds to the $m^2=24$ massive vector field truncation of \cite{Maldacena:2008wh}.

\subsection{Non-supersymmetric truncations}

All the truncations we have listed so far have field content which fills the bosonic sector of AdS$_5$ supermultiplets and so are presumably supersymmetric truncations. It is also useful to consider truncations which contain dynamical fields belonging to different supermultiplets, without keeping the entire multiplet. In this sense these truncations are not supersymmetric, although they are perfectly consistent truncations and solutions of the ten-dimensional equations of motion.  For these truncations, we start with the complete
Lagrangian given in (\ref{eq:lag0}) and (\ref{eq:lag1}).

\subsubsection{Massive vector field}\label{sec:vec}

The first non-supersymmetric truncation we will discuss involves keeping the $m^2=8$ vector field, $b_1^i$, and has already been noted in \cite{Maldacena:2008wh}. The field content in this truncation consists of one component of $b_1^i$ (denoted $b_1$), $\tau_2$, $\rho$, $\sigma$ and $g_{\mu\nu}$.  Note that the graviphoton is turned off here so that even at the lowest level this cannot be supersymmetric. Furthermore, by keeping only one component of $b_1^i$, the $\tau$ equation of motion demands that we must set $\tau_1 = 0$. With this field content, the $D=10$ constraints (\ref{eq:cons}) are trivially satisfied with $\phi_0 = 0$ and $p_2 = 0$, and the Lagrangian (\ref{eq:lag0}) becomes \cite{Maldacena:2008wh}
\begin{eqnarray}
\mathcal L&=&R*1+(24e^{-\fft{16}3\rho-\sigma}-4e^{-\fft{16}3\rho-6\sigma}-8e^{-\fft{40}3\rho})*1
-\ft{40}3d\rho\wedge*d\rho - 5 d\sigma\wedge*d\sigma \nn \\
&&-\ft1{2\tau_2^2}d\tau_2\wedge*d\tau_2 - \ft12\tau_2e^{\fft43\rho+4\sigma}db_1\wedge*db_1 - 4\tau_2e^{-4\rho-2\sigma}b_1\wedge*b_1.
\end{eqnarray}

\subsubsection{Complex massive anti-symmetric tensor}

We can also truncate to theories containing the $m^2 = 9$ complex anti-symmetric tensor field $q_2$. The field content here is given by, $q_2$, $\mathbb A_1$, $B$, $C$, $\tau$, $g_{\mu\nu}$ and $A_1.$ The $D=10$ constraints become $\phi_0 = 0$ and $p_2 = -dA_1 - \ft14d\mathbb A_1$. All the other equations of motion are either satisfied by setting the rest of the fields to zero or can be derived from the Lagrangian
\begin{eqnarray}
\mathcal L&=&R*1+(24e^{-\fft{16}3\rho-\sigma}-4e^{-\fft{16}3\rho-6\sigma}-8e^{-\fft{40}3\rho})*1
-\ft{40}3d\rho\wedge*d\rho - 5 d\sigma\wedge*d\sigma \nn \\
&&-\ft12e^{\fft{16}3\rho-4\sigma}F_2\wedge*F_2 - e^{-\fft83\rho+2\sigma}(p_2\wedge*p_2+4q_2\wedge*\bar q_2) - \ft1{2\tau_2^2}d\tau\wedge*d\bar\tau  \nn \\
&& - \ft12e^{-8\rho-4\sigma}\mathbb A_1\wedge*\mathbb A_1 + \ft{2i}3(q_2\wedge d\bar q_2 - \bar q_2\wedge dq_2) - A_1\wedge p_2\wedge p_2 - 4A_1\wedge q_2\wedge\bar q_2.
\end{eqnarray}
Note that it is consistent to further truncate to a constant axi-dilaton
$\tau = \tau_0$.

\subsubsection{Real massive anti-symmetric tensor}

Along similar lines to the case above for a massive vector field, we can set $A_1=0$ and make a truncation including the $m^2=16$ real anti-symmetric tensor doublet $b_2^i$ by keeping only the graviton coupled to $b_2^i$, $\tau$, $\rho$ and $\sigma.$ Again, the equations of motion for the other fields are trivially satisfied and the constraints are also trivial $\phi_0 = 0$ and $p_2 = 0$. This leaves the Lagrangian
\begin{eqnarray}
\mathcal L&=&R*1+(24e^{-\fft{16}3\rho-\sigma}-4e^{-\fft{16}3\rho-6\sigma}-8e^{-\fft{40}3\rho})*1
-\ft{40}3d\rho\wedge*d\rho - 5 d\sigma\wedge*d\sigma \nn \\
&& - \ft1{2\tau_2^2}d\tau\wedge*d\bar\tau -\ft12 e^{\fft{20}3\rho}\mathcal M_{ij}db_2^i\wedge*db_2^j + 2\epsilon_{ij}b_2^i\wedge*db_2^j.
\end{eqnarray}
As in the previous truncation it is consistent to further truncate to
$\tau = \tau_0$.

\section{Discussion}
\label{sec:discuss}

In the above, we have examined massive reductions of 10-dimensional IIB
supergravity on Sasaki-Einstein manifolds. By utilizing the structure of
SE$_5$, we have given a general decomposition of the IIB fields based on the
invariant tensors associated with the internal manifold. The field content
obtained in five-dimensions completes the bosonic sector of various AdS$_5$
supermultiplets, and in particular they fill out the lowest Kaluza-Klein tower
up to the breathing mode supermultiplet.  This proves, at least at the level
of the bosonic fields, the conjecture raised in \cite{Liu:2000gk} that a
consistent massive truncation may be obtained by truncating to the singlet
sector on the Kahler-Einstein base $B$ (which is $CP^2$ for the squashed
$S^5$) and further restricting to the level of the breathing mode multiplet
and below.

As suggested at the end of Section~\ref{sec:linkk}, it is this truncation
to constant modes on the base $B$ that ensures the consistency of the
reduction.  In a sense, this is a generalization of the old consistency
criterion of restricting to singlets of the internal isometry group,
except that here restricting to singlets of an appropriate subgroup turned
out to be sufficient.  For this reason, we believe it is not that
the breathing mode is special in itself which allows for a consistent
truncation retaining its supermultiplet, but rather that in the examples
given here and in \cite{Gauntlett:2009zw}, the breathing mode superpartners
so happen to be the lowest harmonics in their respective Kaluza-Klein
towers.  It is an unusual feature of Kaluza-Klein compactifications on
curved internal spaces that states originating from different levels of the
harmonic expansion may combine into a single supermultiplet.  Thus, while
the breathing mode is always the lowest state in its tower (being a constant
mode on the internal space), its superpartners may carry excitations on
the internal space.  This does not occur for the $\mathcal N=2$
compactification of IIB supergravity on $SE_5$ (nor does it for $D=11$
supergravity on $SE_7$).  However, in extended theories, such as IIB
supergravity on the round $S^5$, the superpartners will involve non-trivial
harmonics.  In particular, the $\mathcal N=8$ superpartners to the
breathing mode include a massive spin-2 excitation of the graviton involving
the second harmonic (d-waves) on the sphere.  Thus we believe it to be
unlikely that an $\mathcal N=8$ massive truncation with the breathing mode
multiplet will be consistent.

Consistent truncations of the type discussed here have recently been of particular interest in the growing literature on AdS/CFT applications to condensed matter systems. Until recently a strictly phenomenological approach has been taken in this area. In these systems the inclusion of a scalar condensate is required in the gravity theory to source an operator whose expectation value acts as an order parameter describing superconductor/superfluid phase transitions in the strongly coupled system. In the phenomenological approach, the origin of this scalar and its properties have not been of immediate interest; rather the general behavior was determined and many interesting similarities to real condensed matter systems have been noted. However, this approach lacks strong theoretical control in that systems are described by a set of free parameters which can be tuned to provide the property of interest. Recently there has been some work to embed these models in UV complete theories, where the parameters are no longer free but are determined by the underlying features of the theory, such as an origin in string theory. The discussion here has put these reductions into a more general framework and gives further examples of UV complete systems whose duals may have useful applications in the AdS/CMT correspondence.

Given that the fields in these truncations fall into specific supermultiplets it is an obvious and relevant question to discuss their fermionic partners. This would involve reducing the supersymmetry variations and fermion equations in ten-dimensions down to five-dimensions and determining the complete supersymmetric action of these truncations. This is also of interest in terms of AdS/CMT where there has been much interest in describing fermion behavior in condensed matter systems such as the Fermi-liquid theory using the holographic correspondence. In particular, the full supersymmetric action could give us examples of specific interactions studied in these systems coupling scalar condensates to the fermionic excitations
\cite{Chen:2009pt,Faulkner:2009am,Gubser:2009dt}.
We leave the study of the fermionic modes and connections to condensed matter systems to future work.

\begin{acknowledgments}

We wish to thank I. Bah, M. Duff, J. Gauntlett and D. Vaman for stimulating
discussions.
This work was supported in part by the US Department of Energy under grant
DE-FG02-95ER40899.

\end{acknowledgments}


\end{document}